\input stromlo

\title High-$z$ objects and cold-dark-matter cosmogonies: the case of 53W091.

\shorttitle 53W091 and CDM cosmogonies

\author A. Kashlinsky$^{1,2}$ and R. Jimenez$^3$

\shortauthor Kashlinsky \&Jimenez

\affil $^1$ NORDITA, Blegdamsvej 17, Copenhagen DK-2100, Denmark

\affil $^2$ Theoretical Astrophysics Center, Juliane Maries Vej 30, Copenhagen DK-2100 Denmark

\affil$^3$ Royal Observatory, Blackford Hill, Edinburgh EH9 3HJ, UK

\abstract The recently
discovered high redshift galaxy, 53W091, with accurate age measurements 
(Dunlop et al 1996) provides a measure of the
small-scale power of the primordial density field and, as we show, a crucial
test of the inflation-inspired models. It allows $\Omega$=1 cosmologies 
only for low values of $H_0$, but then pushes formation of that
galaxy to redshift much greater than allowed for by the cold-dark-matter
density field. 
Cold-dark-matter (CDM) models with cosmological constant ($\Lambda$)
and low $\Omega$ would decrease the redshift at which this galaxy has collapsed.
However, in CDM models decreasing $\Omega$ suppresses the small
scale power in the density field and 
this effect turns out to be dominant. We estimate the mass of  the galaxy 
and show that it represents a very rare and unlikely
event in the density field of such models. 
Similar problems would occur in other 
modifications of the CDM cosmogonies.

\section Introduction

Inflation provides an attractive framework to explain the origin of the
large scale structure in the Universe. At the same time it makes 
two firm predictions: 1) if started from inhomogeneous
initial conditions and later homogeneised by inflationary expansion, the Universe 
must be flat today to be consistent with the microwave background anisotropy measurements
(Kashlinsky, Tkachev $\&$ Frieman 1994); and 2) the primordial power spectrum of the 
density field is of the Harrison-Zeldovich form modified only by a different growth
rate for sub- and super- horizon fluctuations during the radiation dominated era
(e.g. Bardeen et al 1986). 

The epoch of galaxy formation reflects the small scale power in the primordial
density field as well as cosmological parameters and thus provides an additional test 
of cosmological theories. 

We show that the high-$z$ galaxy, 53W091, recently discovered 
by Dunlop et al (1996; D96), provides a sensitive test of
inflationary models (Kashlinsky $\&$ Jimenez 1996).
The age of the galaxy estimated from its red stellar population formally
rules out Einstein-de Sitter for $H_0$$>$ 50 km/sec/Mpc. For smaller values
of $H_0$ the redshift of formation of the galaxy is so high that 53W091 will have to be a
$>$10-$\sigma$ event for the density field predicted by standard 
inflationary paradigm. To reconcile such old and as we show below massive galaxy 
with inflationary paradigm 
requiring flat Universe, the alternative may seem at first to assume a
cosmological constant, $\Lambda\equiv 3H_0^2 \lambda$, dominated 
Universe with $\Omega+\lambda$=1. Inter alia this model was proposed to account for
the excess large-scale power seen in galaxy catalogs within the 
inflationary picture (Efstathiou et al 1990). This would also allow
more time for the galaxy to evolve and push its formation to lower $z$. However,
increasing the cosmological constant, $\lambda$,
and decreasing $\Omega$ would
at the same time suppress the small scale power in the density field and push
galaxy collapse and formation to more recent epochs. We show that the latter makes 
53W091 a $>$(6-7)-$\sigma$ event in such models. Consequently the galaxy represents
a challenge to all inflation inspired cosmogonies.

\section Parameters of 53W091

The discovery of and data on 53W091 were recently presented in D96.
Its red color is indicative of an old stellar population and its
blue apparent magnitude $V$=26
implies a large luminous mass.  They have been able to identify
late-type stellar
absorption features in the spectrum of the galaxy which allow one to
determine the age.
The spectrum for 53W091 was obtained in the range 2000 to 3500 \AA,
where the main contribution to the integrated light comes from the
main sequence stars.
Hence D96 built a series of synthetic
spectra at different ages with the main contribution coming from the
main sequence stars. This was done
using the set of stellar atmospheres models from Kurucz (1992) and a grid
of stellar interior models from Jimenez \& MacDonald (1997). 

The best fit was found for the age of $t_{age}$=3.5Gyr
(D96).  An independent evidence that the age of
53W091 cannot be less than 3 Gyr comes from transition
breaks in the spectrum.
The two of them, at 2600 and 2900 \AA ,
 were computed for different metallicities (1/5 $Z_\odot$,
$Z_\odot$, and 2 $Z_\odot$);
the observed amplitudes showed that the breaks
cannot be reproduced if $t_{age}$$<$3Gyr (D96).
The numbers
for $t_{age}$ were computed in D96 assuming 53W091 to be an
elliptical, i.e. assuming that the $\alpha$-nuclei elements
are enhanced with respect to the Sun with
the typical enhancement factors being [$\alpha$/Fe]=0.3-0.5.
We therefore follow D96 and adopt for the present discussion
the age uncertainty of the 53W091 galaxy to be no more than 0.5 Gyr with
the most likely value of $t_{age}$=3.5Gyr. Pushing the age to the lower
limit of the range, $t_{age}$=3Gyr, would at the same time require high
metallicity ($Z\geq$2 $Z_\odot$) typical of nuclei
of ellipticals. (The galaxy was observed with an aperture of 4 '' making the
nucleus region unresolved).

One needs to estimate the mass of 53W091 in order to test its implications
for galaxy formation. To compute the total {luminous} mass
of the galaxy we used the integrated synthetic spectra from D96. 
The flux was then scaled
assuming the Miller-Scalo IMF until it matched the observed fluxes in all,
V,J,H,K, bands from D96. The effect
of changing the slope of the IMF on the total luminous mass is generally
small: $\sim$1\% when the IMF slope, $\alpha$, changes from 2.5 to 3.5.
We computed the total mass in stars in the galaxy for different cosmologies.
For the case of $\Omega=1.0, \Lambda=0, H_{0}=60$  km/sec/Mpc, the 
mass was calculated for 3 different metallicities (1/5, 1 and 2 $Z_\odot$). We found a value for 
the mass of 53W091 between $0.8 \times 10^{12}$ and $1.3 \times 10^{12}$ $M_\odot$. 
Subsequently, 
we studied models with solar metallicity and $\Lambda=0, H_0=60$  km/sec/Mpc but 
for different values of $\Omega$ (0.2, 0.3, 0.4). In this case the computed 
mass was found between $1.1 \times 10^{12}$ and $1.9 \times 10^{12}$ $M_\odot$. Finally, 
we computed the mass for different values of $H_0$ with $\Omega=1$ and solar 
metallicity. The mass values are in between  $1.1 \times 10^{12}$ and 
$2.4 \times 10^{12}$ $M_\odot$ (see Kashlinsky \& Jimenez (1996) for more details).
Because of the increasing distances, 
for flat $\Omega$+$\lambda$=1 models the mass is even
higher. E.g. for the flat $\Lambda$-dominated Universe 
with $\Omega$=0.2 the total stellar mass for
53W091 is $1.8 \times 10^{12} M_\odot$. The trends with $H_0$
and $\Omega$ are very similar to the $\Lambda$=0
case; for brevity we do not present the numbers here.
In order to account for different rates of star formation we
adopted the star formation prescription described in Chambers \& Charlot 
(1990) and computed a series of integrated
synthetic spectra with different values for the typical star formation rate
parameter $\tau$.
The change in the total star mass for different star formation laws
is small, particularly in the likely case of $\tau$$<$2Gyr.
Larger $\tau$ lead to $Z \gg Z_\odot$
and also give larger $M$: e.g. if $\tau$$\geq$3Gyr,the mass estimates
shown in Table 1 {increase} by 37\%. The numbers vary
little with model or cosmological parameters and show that
53W091 has $\geq 10^{12}M_\odot$ in stars alone inside the aperture 
of 4''.

%

\section Cosmology

Following the results from the previous section we assume in what follows that the data 
on 53W091  imply that the galaxy at $z$=1.55 has mass in excess of $10^{12}M_\odot$ 
and its stellar population has age of $\simeq 3.5$ Gyr. What are then the cosmological
implications of at least one object in the Universe having collapsed (formed galaxy) 
on mass scale of $>10^{12}M_\odot$ at least 3.5 Gyr before the redshift of 1.55?

The left box in Fig.1 shows the redshift
$z_{gal}$ at which the galaxy 53W091 must have formed its first stars for
$\Omega$+$\lambda$=1 Universe. 
Solid lines correspond
to $t_{age}=3$Gyr, dotted to 3.5 Gyr and dashed to 4 Gyr. Three types of
each line correspond to
$H_0$=60,80 and 100 km/sec/Mpc. 
One can see that
the value of $z_{gal}$ decreases as {both} $\Omega$ and $h$ decrease.
On the other hand, in the low-$\Omega$ CDM cosmogonies the
small-scale power is
also reduced as the {product} $\Omega h$ decreases;
this would at the same time delay collapse of first galaxies until
progressively smaller $z$.

\figureps[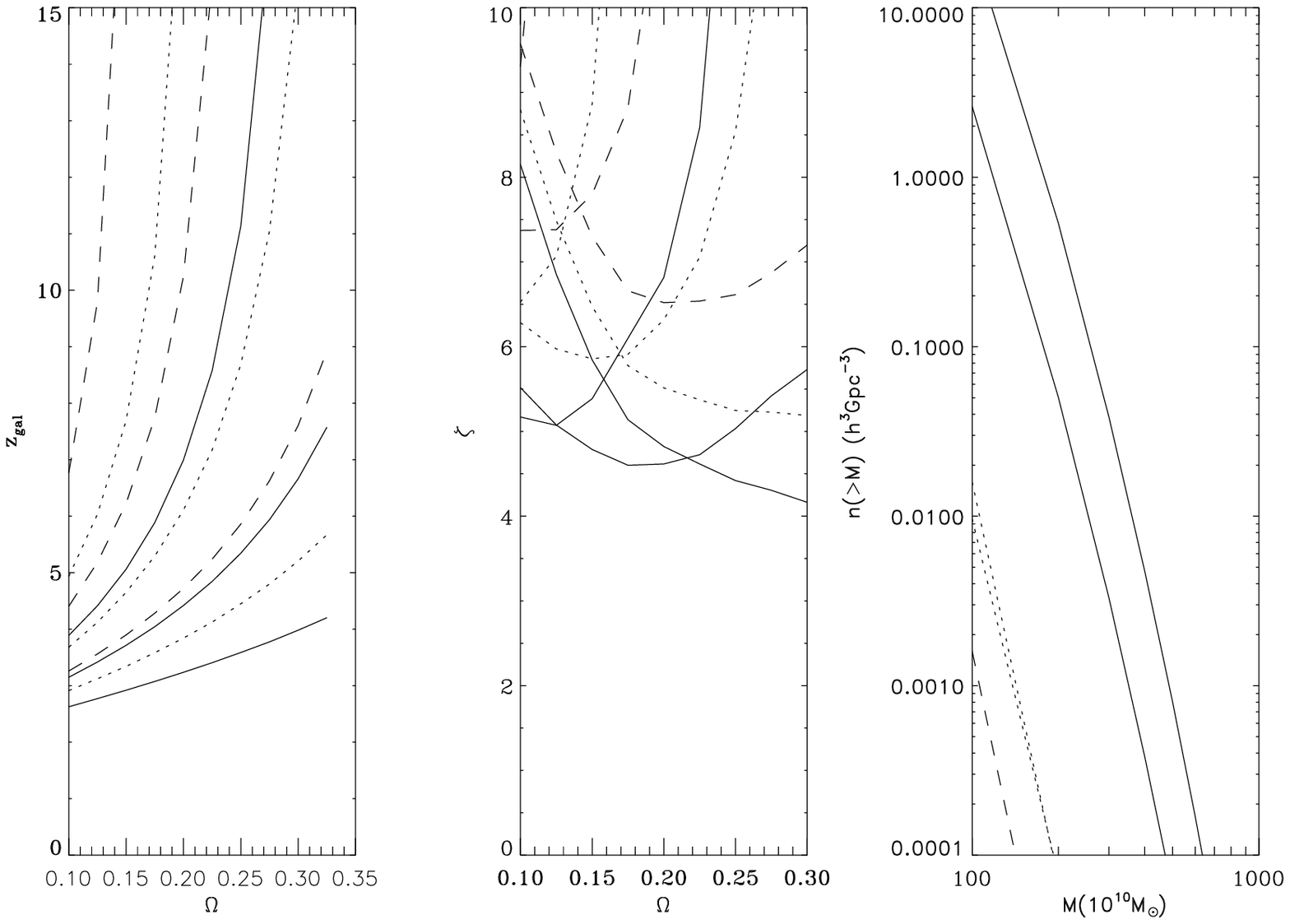,.4\vsize] 1. 

To quantify this we proceed in the manner outlined in Kashlinsky (1993). This involves
the following steps (see Kashlinsky $\&$ Jimenez 1996 for details): 1) Specify
the primordial power spectrum, $P(k)$, of the density field at some initial redshift
$z_i \gg$1 when the density field is linear on all scales. The power spectrum depends
on the initial power spectrum, assumed to be Harrison-Zeldovich, and the transfer function
which accounts for the evolution of the shape of the power spectrum in the linear regime.
The latter depends in such models only on the product $\Omega h$ and
was adopted from Bardeen et al (1986).
2) Compute the amplitude,
$\Delta_8$, of
that field at $z_i$ on the scale of $8h^{-1}$Mpc that produces the observed unity 
rms fluctuation in galaxy counts today, or a $1/b$ amplitude in mass fluctuation ($b$
is the bias factor) at $z$=0; 3) Compute the density, $\delta_{col}(z)$, the fluctuation had to have 
at $z_i$ in order to collapse at $z$. 4) A convenient quantity to describe 2) and 3) is
$Q(z)\equiv \delta_{col}(z)/\Delta_8$. For $\Lambda$-dominated flat Universe and in 
the limit of $1+z_{gal} > \Omega^{-1/3}$ it can be approximated as $Q(z)\simeq 3
\Omega^{0.225} b^{2/3}(1+z)$. 5) $b$ is determined by normalizing the density
distribution given by $P(k)$ to the COBE-DMR maps (Bennett et al 1994; Stompor et al
1995). 6) Given $P(k)$ we compute the rms fluctuation, $\Delta(M)$, over a region containing
mass $M$. 7) The quantity $\zeta \equiv Q(z_{gal}) \Delta_8/ \Delta(M)$ then describes the 
number of standard deviations an object of mass $M$ had to be in order to collapse at $z_{gal}$
in the cosmological model specified by $P(k)$.

 The values of $\zeta$ for $M$=$10^{12} M_\odot$ are plotted versus $\Omega$ in 
the middle box of Fig.1 for $\Omega$+$\lambda$=1 for various
values of $t_{age}$ and $h$. As in the left box solid lines correspond to
$t_{age}$=3Gyr, dotted to 3.5 and dashes to 4 Gyr. Three types of each line
correspond to $h$=0.6, 0.8 and 1 going from bottom to top at large values
of $\Omega$ at the right end of the graph. The line for $t_{age}$=4Gyr and
$h$=1 lies above the box. As the figure shows, this galaxy must
represent an extremely rare fluctuation in the density field specified by the 
low-$\Omega$ flat CDM models.
Note that the total mass of 53W091 must be at
least a factor of 10 larger in which our conclusion will be much stronger. 

The right box shows the expected co-moving number density of such galaxies, $n(>M)$,
in units of ($h^{-1}$Gpc)$^{-3}$ vs the total (dark+luminous) mass for
$t_{age}$=3.5Gyr. It was computed using the Press-Schechter (1974) prescription.
The numbers for the co-moving number density
$n(>M)$ were computed for $\Omega$=0.1,0.2,0.3 and
$h$=1,0.8,0.6. Solid lines correspond to $\Omega$=0.1 and to 
$h$=1,0.8
from top to bottom. Dotted lines correspond to $\Omega=0.2$ and $h=0.8,0.6$ 
from top to bottom respectively. The dashed
line corresponds to $\Omega=0.3$ and $h$=0.6. 
The models not shown lie below
the box. 

\section Conclusions

We have shown that within the framework of the 
standard and $\Lambda$-dominated CDM models objects like 53W091
must be extremely rare in the Universe. There exists a
narrow range of parameters ({total} mass of $10^{12}M_\odot$, age of
$<$ 3 Gyr, $\Omega$=0.1 and $h \geq$0.8) where one expects to find a few of
such
objects with each horizon, $R_{hor} \simeq 6 h^{-1}$Gpc for $\Omega$=1, 
but for most
cases the number density of such objects is less than one per horizon volume.
The data suggest that the Universe contains more collapsed 
galaxies at early times than the modified CDM models would predict. 
Such objects can be
most readily accommodated by assuming both 1) low $\Omega$
Universe and 2) the small-scale power in the primordial power spectrum
in excess of that given by simple inflationary models. E.g.  the
necessarily high redshift of galaxy formation can be produced
in string models (Mahonen, Hara \& Miyoshi 1995) or in the phenomenological
primeval baryon isocurvature model (Peebles 1987).

\references

 Bardeen, J.M., Bond, J.R., Kaiser, N., Szalay, A.S. 1986, Ap.J., {\bf
304}, 15.

 Bennett, C. et al. 1996, Ap.J.,{\bf 464}, L1.

 Dunlop, J, Peacock, Spinrad, H., J., Dey, A., Jimenez, R. Stern, D. and
Windhorst, R. 1996, Nature, {\bf 381}, 581. (D96)

 Jimenez and MacDonald 1997, {\it in preparation}.

 Kashlinsky, A. 1993, Ap.J., {\bf 406}, L1.

Kashlinsky, A. and Jimenez, R., 1996, Ap.J.Letters, in press.

 Kashlinsky, A., Tkachev, I. and Frieman, J. 1994, Phys.Rev., {\bf
73}, 1582.

 Kurucz, R. ATLAS9 Stellar Atmosphere Programs and 2km/s Grid CDROM Vol. 13
(Smithsonian Astrophysical Observatory, Cambridge, MA, 1992).

 Mahonen, P., Hara, T. and Miyoshi, S.J. 1995, Ap.J., {\bf 
454}, L81.

 Peebles, P.J.E. 1987, Nature, {\bf 327}, 210.

 Press and Schechter 1974, Ap.J., {\bf 187}, 425.

 Stompor, R., Gorski, K.M., Banday, A.J. 1995,MNRAS, {\bf 277}, 1225.

\vfill
\eject

QUESTION: 
Josh Barnes: Could this galaxy be lensed and thus have amplified flux?

ANSWER: 
In principle yes, but this is not very likely. A way out  for
CDM models could be to assume that
the stars in 53W091 galaxy formed in smaller collapsed objects and
later merged into the galaxy as we see it.
However, this too appears unlikely. 
53W091 looks like a normal old relaxed giant elliptical. The
physical
diameter size corresponding to the aperture of 4`` at
$z$=1.55 is 29, 20 and 17 $h^{-1}$Kpc for $(\Omega,\lambda)$=
(0.1,0.9),(0.1,0) and (1,0) respectively; this is typical of diameters of giant
ellipticals. It would
be hard to see how mergers could have led to the substructure dissipating its
energy to collapse to what appears a normal elliptical radius in such a short
time. The galaxy has very red colors, indicating no star formation for
the past 3Gyr, so this would further have to occur without significant star
formation expected from merger events. 
Indeed,  if the galaxy had undergone mergers
3.5 Gyr ago, the following star formation events would have created a
significant blue
component in its spectrum contrary to what is observed.
Thus it is likely that for at least 3.5 Gyr  53W091 did not undergo any major
merging
events and its mass 3.5 Gyr ago had to be the same. Even invoking mergers would
not help much. At the relevant masses and the CDM power spectra 
$\zeta$$\propto$$M^{0.1-0.15}$ for the relevant cosmological
parameters, so in order to decrease $\zeta$ appreciably 53W091 must have 
undergone a very large number of mergers in a short time. The $\zeta-M$ 
dependence would also mean that even if lensing could have led us to overestimate
its true luminosity and mass, this would not help much in trying to account
for 53W091 in the framework of CDM models.

QUESTION: 
Ben Dorman: This is a cautionary note about drawing inferences from 2600/2900 \AA 
 break indices based on the Kurucz fluxes. These model spectra may not reproduce
the spectra of the real stars well enough to be sure of the results. There are not
yet enough empirical spectra in the wavelength range to check the calculations.
It is true, however, that the continuum level observed implies, for a wide range in
metallicity, an age of $\sim 3$ Gyrs, but there is still room for a younger interpretation
if $Z$ is $>$ 2 solar. (Though one might argue that luminosity weighted mean abundances
much higher than this are unlikely).

ANSWER:
It is true that Kurucz models are mostly theoretical. 
Nevertheless, there are several points that 
support the Dunlop et al age values for 53W091: 
1) An age of $\sim$ 3 Gyr was also obtained using the Bruzual and 
Charlot code which is based on the IUE observed spectra.  Analyzing the 
breaks values for single stars from the 
IUE atlas shows that there are stars with breaks values 
like those of 53W091 and some with higher values. The stars with breaks 
similar to 53W091 are F6 class with an age of $\sim$ 3.5 Gyr.
2) The breaks are NOT so sensitive to metallicity (see Dunlop et 
al 1996, Nature 381, 581). As mentioned in the question the mean 
metallicities of comparably massive giant ellipticals at low redshift are 
at most only mildly super-solar when averaged out to the extension of the 
galaxy. 3) The 2600/2900 \AA breaks from Kurucz models give  a good fit 
to the age of the Sun. In case these breaks were modified (decreasing them to 
produce a younger age for 53W091) they would yield a MS turnoff age 
of $<$ 3 Gyr for the Sun.

\bye